\documentclass[12pt]{article}
\begin{document}

\title{NEUTRINO DATA AND NEUTRINO-ANTINEUTRINO TRANSITION}

\author{E.N. Alexeyev and L.V. Volkova \\
        \it Institute for Nuclear Research of Russian Academy of Sciences \\
        \it 60th October Anniversary pr.7a, 117312, Moscow, Russia}

\date{April 27, 2004}

\maketitle

\begin{abstract}
A  problem,   whether  a  neutrino-antineutrino  transition  could  be
responsible  for  the  muon  neutrino  deficit  found  in  underground
experiments (Super-Kamiokande, MACRO, Soudan 2) and in the accelerator
long-baseline  K2K  experiment,  is  discussed  in   this  paper.  The
intention of the work is not consideration of concrete models for muon
neutrino -antineutrino transition but a desire to attract an attention
to another possibility of understanding the  nature  of  the  measured
muon neutrino deficit in neutrino experiments.
\end{abstract}

\section{Introduction}

The deficit of up going muon neutrinos in comparison with the expected
ones was  found  in different underground experiments \cite{c1,c2,c3}.
In more detail and  for wide range of energy this effect  was registed
at the Super-Kamiokande detector  \cite{c1}.  Some later the effect of
the  muon  neutrino  disapearance  was  cofirmed  in  the  accelerator
long-baseline K2K  experiment  \cite{c4}. Results of these experiments
were  interpreted  as the  existence  of  muon  neutrino-tau  neutrino
oscillations. The  corresponding values of oscillation parameters were
calculated.

It is necessary to note that the discussed experiments are experiments
on the  disapperance of initial  neutrinos, in which were not reliably
defined types of newborn neutrinos.

\section{Neutrino-antineutrino transition}

Another possibility can be considered in order to explain the observed
muon neutrino deficit. Let us assume that muon  neutrinos transit (not
oscillate!)  to  muon  antineutrinos.  As  a  cross  section  of  muon
antineutrino-nucleon interaction  is  less  than muon neutrino-nucleon
one (about  2 times at  energy of  about 1GeV), a  number of  neutrino
events detected will be less  than  the expected one. We can  estimate
the ratio value ($R$) of the number of muons (produced  by atmospheric
neutrinos deep  underground) arised in  the case of realizing the muon
neutrino-muon antineutrino transition to that with no transition.

Fluxes of  atmospheric neutrinos coming  to the sea level at different
directions for  a wide energy range  were calculated in  \cite{c5}. In
\cite{c6} calculations were made to study the main  features of fluxes
of  atmospheric  muon   neutrinos   responsible  for  a  prodution  of
unerground muons at, in  particular,  middle and effective energies of
neutrinos responsible  for  these  muons,  neutrinos  to antineutrinos
ratio. Our estimations are based on these calculations.

The ratio value  ($R$) estimated under  conditions that all  up  going
muon neutrino transit to muon antineutrino are equal to as follows
\begin{eqnarray}
& 0.625 &  \mbox{ for } E_{\nu_\mu} > 0.3 \mbox{ GeV} \\                                                       
& 0.585 &  \mbox{ for } E_{\nu_\mu} > 10 \mbox{ GeV}
\end{eqnarray}

As it is seen from these  values, the ratio R is aproximately the same
paramerter for wide energy range.

The $R$-estimations obtained can be  compared  with  the  experimental
data from Super-Kamiokande published in [1]. There is the double ratio
of events experimentally measured to  ones  calculated  with the Monte
Carlo method in this work
\begin{equation}
R = \biggl(\frac{\mu}{e}\biggr)_{DATA} / \biggl(\frac{\mu}{e}\biggr)_{MC}
\end{equation}
where $(\mu/e)$ --- $\mu$-like to $e$-like events ratio

For two  energy ranges  of $E  < 1.33$  GeV and $E > 1.33$  GeV $R$  -
values are the following
\begin{eqnarray}
&& R_{sub-GeV} = 0.658 \pm 0.016 \pm 0.055 \\                                                                                                   
&& R_{multi-GeV+PC} = 0.702^{+0.032}_{-0.030} \pm 0.101
\end{eqnarray}
As  the  number  of  experimentally  measured  e-like events in  Super
Kamiokande coincides  with the expected  one, it is possible to assume
with reasonable accuracy that
\begin{equation}
R = \biggl(\frac{\mu}{e}\biggr)_{DATA} / \biggl(\frac{\mu}{e}\biggr)_{M}
\approx \mu_{DATA} / \mu_{MK}
\end{equation}
Comparing the estimation values (1)  and  (2)  with the experimentally
measured values (4) and (5) from Super Kamiokande, one can arrive to a
conclusion that  the  hypothesis  of  muon  neutrino-muon antineutrino
transition is  in a good agreement  with the experimental  data within
the shown errors.

\section{Conclusion}

The  alternative  variant  of  an  explanation  of  the  observed muon
neutrino  deficit  is suggested in this paper.  Namely,  the  possible
transition  of  muon  neutrinos  to  muon  antineutrinos  during  muon
neutrino path through  the  Earth was  used  to obtain the  estimation
values for comparing  with the experimental Super Kamiokande data. The
result was found to be positive within experimental errors. This means
that the hypothesis of muon neutrino-muon antineutrino transitions can
not be ruled out and it would be very desirable to test it.

Both the  current experiments and  the future ones are very expensive,
it seems very important to foresee the possibility of verifications of
above mentioned  hypothesis  in  these  experiments.  For instance, it
would  be  most  importantly  to install a magnetic  spectrometer  for
charge  mesurments  of neutrino-produced  muons  in  the  current  K2K
experiment  in  the  place  of the Super Kamiokande  disposition.  The
presence of positively  charged muons in a magnetic spectrometer would
be  the  direct  corroboration  of muon neutrino transitions  to  muon
antineutrinos.

\end{document}